# Hunting Tomorrow's Leaders: Using Machine Learning to Forecast S&P 500 Additions & Removal

Vidhi Agrawal[1], Eesha Khalid[2], Tianyu Tan[3], Doris Xu[4]

Columbia University

New York, NY, United States

va2504@columbia.edu[1], ek3365@columbia.edu[2], tt2976@columbia.edu[3], jx2577@columbia.edu[4]

*Abstract*— This study applies machine learning to predict S&P 500 membership changes—key events that profoundly impact investor behavior and market dynamics. Quarterly data from WRDS datasets (2013 onwards) was used, incorporating features such as industry classification, financial data, market data, and corporate governance indicators. Using a Random Forest model, we achieved a test F1-score of 0.85, outperforming logistic regression and SVC models. This research not only showcases the power of machine learning for financial forecasting but also emphasizes model transparency through SHAP analysis and feature engineering. The model's real-world applicability is demonstrated with predicted changes for Q3 2023, such as the addition of Uber (UBER) and the removal of SolarEdge Technologies (SEDG). By incorporating these predictions into a trading strategy—buying stocks announced for addition and shorting those marked for removal—we anticipate capturing alpha and enhancing investment decision-making, offering valuable insights into index dynamics.

*Keywords—S&P 500, Index Inclusion, Prediction, Machine Learning, Random Forest, SVC, Logistic Regression, SHAP*

## I. INTRODUCTION

The S&P 500 index, a critical benchmark for the U.S. equity market, significantly influences investor behavior and portfolio strategies. Passive investing, which tracks indices like the S&P 500, accounts for 20-30% of the value of U.S. equities, amplifying the importance of changes to the index. When a company is added to the index, its stock price typically rises due to anticipated demand from index fund managers and speculative trading [8]. For instance, Tesla's inclusion in the S&P 500 in December 2020 led to extraordinary trading activity and price performance, underscoring the significance of such changes. Conversely, deletions often lead to price declines as portfolios are adjusted to reflect the updated index composition.

Between December 13, 2019, and September 24, 2024, we analyzed the effects of S&P 500 additions and deletions using announcements from S&P's website and data from CRSP. Our findings show that stocks added to the index experienced significant price surges, while those removed faced declines, reflecting the predictable impact of index fund adjustments [1]. However, outliers like Ingersoll-Rand (IR), which underwent a ticker change to TT after a spinoff, and Apartment Investment and Management Co. (AIV), affected by an unadjusted stock split, introduced distortions. Excluding these outliers provided a clearer understanding of the systematic impacts of index changes.

The 1-day, 2-day, and 7-day price movements following S&P 500 announcements (excluding IR and AIV) highlight clear alpha capture opportunities.

TABLE I. STATISTICS FOR ADDITION AND REMOVAL ACTIONS

```
Statistics for Addition Actions (excluding IR):
        price_1d_move (%)  price_2d_move (%)  price_7d_move (%)
mean            2.855223           2.690931           3.099722
min            -3.719058          -5.252245         -11.230372
max            17.670588          20.762492          27.876204
median          1.634824           1.372396           1.350281
Statistics for Deletion Actions (excluding AIV):
        price_1d_move (%)  price_2d_move (%)  price_7d_move (%)
mean           -0.168411          -0.156093          -2.189437
min           -14.220992         -10.470677         -15.457413
max             7.830343          13.368580           6.492718
median         -0.265913          -0.448708          -2.060282
```

Stocks announced for addition saw significant price increases due to higher demand, supporting a long strategy, while those marked for removal declined due to reduced demand and selling pressure, favoring a short strategy. Building on these insights, we turn to machine learning to predict future S&P 500 additions and deletions [2], [10]. By leveraging historical patterns and financial indicators, we aim to create models that identify potential changes early, allowing investors to capture alpha with greater precision.

## II. DATA

### A. Data Sources and Extraction Process

The dataset used for this study was constructed by combining data from multiple sources, specifically the Wharton Research Data Services (WRDS) platform. The data spans from January 1, 2013, to December 31, 2023, accounting for the financial and operational characteristics of firms and their potential inclusion in or exclusion from the S&P 500 index [9].

1. **CRSP Data**:

Data from the CRSP database was used to capture daily stock prices, market capitalization, and trading volumes of firms. A query was designed to extract data for firms with valid daily market capitalization values. Key variables include: Daily market capitalization, stock price, trading volume, and permno (unique identifier for companies).

The data was filtered to include the last trading day of each quarter. Firms were ranked by market capitalization, and the top 800 firms for each quarter were retained for further analysis.

2. **Compustat Fundamentals**:

Financial statement data from Compustat was integrated to capture firm-level financial indicators, including: Total assets, total liabilities, net income, operating income, cash flow from operations, ratio of current assets to current liabilities, ratio of total liabilities to equity, return on assets, return on equity, earnings per share, and book value per share [3].
Lagged values for these metrics were computed to ensure temporal alignment with the prediction task. Data cleaning steps, including the use of coalesce, were applied to handle missing values.

3. **IBES Data**:

Analyst coverage data from IBES was utilized, focusing on the number of analysts providing earnings estimates for each firm. This variable (num_analysts_covering) provides insight into the level of market attention a firm receives.

4. **Audit Analytics**:

Data on auditor changes and financial restatements was sourced from Audit Analytics: Number of auditor changes for each firm and number of financial restatements.

5. **S&P 500 Membership Data**:

Data on S&P 500 membership was retrieved from CRSP's

dsp500 table. This dataset includes the inclusion and exclusion dates for each firm in the S&P 500. This information was merged with quarterly data, creating a binary label: **in_sp500**. A label indicating whether a firm was included in the S&P 500 at the end of a given quarter (1 for inclusion, 0 otherwise).

*B. Data Integration*

The datasets were merged using unique firm identifiers (permno, gvkey and ticker) and aligned based on temporal dimensions, such as calendar dates and quarters. Linking tables from CRSP and Compustat ensured proper matching of firm-level records across datasets.

*C. Final Dataset*

The final dataset includes the following variables:

- Firm-level financial metrics (e.g., stock price, market capitalization, total assets, ROA, EPS).
- Market performance indicators (e.g., average trading volume over the previous three months, one-month returns).
- Analyst coverage and audit-related data.
- Industry classification (e.g., SIC codes, industry names derived from hierarchical SIC groupings).

The target variable, **in_sp500**, serves as the binary label for predicting S&P 500 inclusion.

This comprehensive dataset enables the application of machine learning techniques to predict quarterly changes in S&P 500 membership based on firm characteristics and market performance.

FIGURE I. CORRELATION HEATMAP OF NUMERIC FEATURES WITH S&P MEMBERSHIP

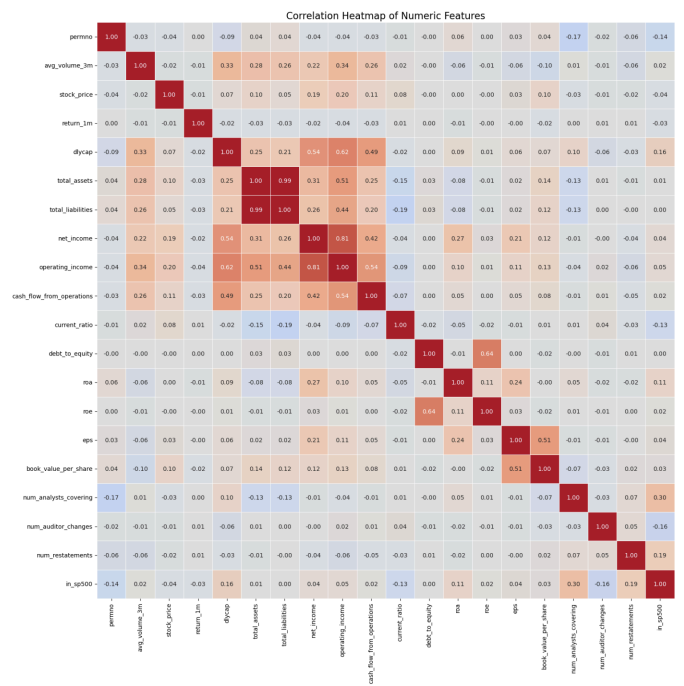

*D. Exploratory Analysis*

1. **Distribution of Financial Metrics**

Density plots for total assets, daily market capitalization, and net income reveal significant skewness in the data, with most firms clustered at lower values for these metrics. Firms in the S&P 500 index (marked as 1) tend to have slightly higher densities at larger values compared to non-S&P 500 firms (0). This suggests that larger firms, in terms of assets, market capitalization, and profitability, are more likely to be included in the S&P 500.
*Total Assets:* Firms in the S&P 500 exhibit a broader distribution toward higher asset values compared to non-S&P 500 firms.

*Daily Market Capitalization:* S&P 500 firms generally have higher market capitalizations, reflecting their prominence in the equity market.
*Net Income*: Positive net income is more prevalent among S&P 500 firms, indicative of their financial stability and profitability.

### 2. Pairwise Relationships

Pairplots between different variables give insights into the relationships between key financial metrics such as total assets, market capitalization, net income, return on assets (ROA), and return on equity (ROE). Notable observations include:

- Strong positive correlation between total assets and market capitalization, with S&P 500 firms occupying the upper range of both metrics. This indicates that these firms are both asset-rich and highly valued by the market. Non-S&P 500 firms are concentrated in the lower range of both axes.

- Firms with higher net income tend to have better ROA and ROE, but this relationship is more pronounced for S&P 500 firms.

- Non-S&P 500 firms are more widely dispersed across lower ranges of financial performance metrics.

FIGURE II. MARKET CAP VS TOTAL ASSETS BY S&P 500 STATUS

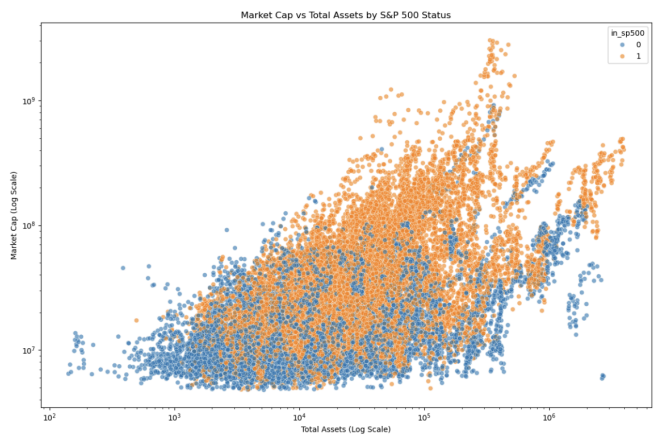

### 3. Industry Distribution

The industry distribution chart underscores the sectoral composition differences between S&P 500 and non-S&P 500 firms:

- S&P 500 membership is heavily concentrated in sectors such as Manufacturing, Technology, and Finance, which are traditionally associated with large-cap companies.

- Non-S&P 500 firms are more evenly distributed across industries but dominate smaller sectors like retail trade and public administration.

- The overrepresentation of manufacturing and technology sectors among S&P 500 members suggests that sectoral dynamics play a critical role in index inclusion.

FIGURE III. INDUSTRY DISTRIBUTION BY S&P 500 STATUS

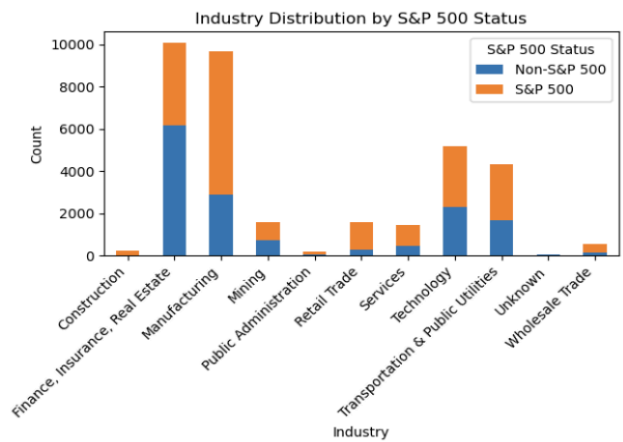

### 4. Hypotheses based on Exploratory Analysis

Based on the exploratory analysis, several hypotheses can be formulated regarding the factors influencing S&P 500 membership:

- Firm Size Hypothesis: Larger firms (in terms of total assets and market capitalization) are more likely to be included in the S&P 500.

- Profitability Hypothesis: Firms with higher net income, ROA, and ROE are more likely to qualify for inclusion due to their financial stability.

- Sectoral Bias Hypothesis: Certain industries, such as manufacturing and technology, are overrepresented in the S&P 500 due to their economic significance and are more likely to be included.

- Market Valuation Hypothesis: A strong correlation exists between market capitalization, number of analysts covering a stock and S&P membership, suggesting that investor perception plays a key role.

These hypotheses will guide further modeling efforts to predict future additions to the S&P 500 index.

## III. MODEL

*A. Data Processing*

The final combined S&P dataset contained a small number of duplicates post-join, which were subsequently removed. Approximately 15-16% of values were missing in certain financial metrics, such as Total Assets, Trading Volume, Current Ratio, and Cash Flow from Operations etc**.** Missing values were imputed using a forward-fill technique, where the missing value was replaced with the rolling mean of the previous two quarters for the respective company (ensuring Point-in-time data). ([Source code](#))

*B. Feature Engineering*

To enhance the predictive power of the model, various derived features were tested. **Market Capitalization (Market Cap)** was excluded from the final model as the analysis focused on identifying factors other than market cap that influence the addition or removal of companies from the S&P index. Key feature engineering steps included:

- *Datetime Features:* Extracting year, month, and day-of-the-week from timestamps.

- *Multicollinearity Handling:* Dropping columns with a correlation coefficient greater than 0.7 to minimize redundancy.

- *Lag Features*: Creating lagged columns to capture financial metrics from the previous quarter, ensuring point-in-time data integrity for predictions.

- *Binary Indicator (Last_Quarter_Positive)*: Introducing a binary variable to indicate whether the company's last quarter earnings were positive, based on the **Earnings Per Share (EPS)** metric.

- *Growth Factor*: Introducing a feature to capture quarter-over-quarter (QoQ) growth in the company's stock price.

*C. ML Model*

The dataset comprised **22,870 records across 23 features**, spanning **42 quarters**. The data was split into:

- Training Set: 35 quarters.
- Validation Set: 5 quarters.
- Test Set: 2 quarters (out-of-sample data for final evaluation).

Numerical features were standardized, and categorical variables, such as Industry and Last_Quarter_Positive, were one-hot encoded to ensure compatibility with machine learning algorithms.

Multiple classification algorithms were tested, including:

- Baseline Logistic Regression
- Random Forest
- Support Vector Classifier (SVC)

*Hyperparameter Tuning*

Based on the initial performance, Random Forest was chosen and tuned to see if it yielded better results. We used GridSearchCV to get the best parameters for n_estimators, max_depth and min_samples_split. The parameters that yielded optimal results were n_estimators = 200, max_depth = None and min_samples_split = 2 in combination with all other default parameters.

TABLE II.    MODEL F1 SCORES

| | Validation_f1_score | Test_f1_score (oos) | Cross-validation score |
|---|---|---|---|
| Random Forest | 0.87 | 0.85 | 0.8 |
| Logistic Regression | 0.771 | 0.7 | 0.72 |
| Support Vector(Class | 0.76 | 0.74 | 0.66 |

Among the three algorithms tested, Random Forest consistently showed the best performance, achieving a validation F1-score of 87% and an out-of-sample test F1-score of 85%. To ensure robustness of the results, we performed cross-validation using TimeSeriesSplit with a K-fold value of 5, to preserve the temporal structure of the data during training and testing.

For further validation of the model's predictive behavior, SHAP (SHapley Additive exPlanations) plots were generated. This game-theory based interpretability method provided insights into how individual features contributed to the model's predictions, allowing us to confirm our hypotheses regarding feature importance and their impact on the target variable [6].

FIGURE IV.  SHAP SUMMARY PLOT- RANDOM FOREST

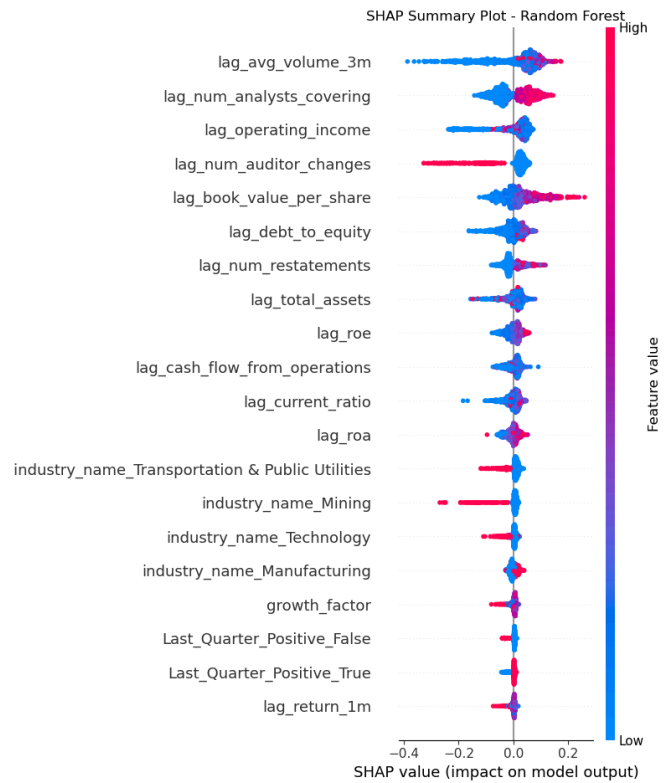

## IV. RESULTS

The machine learning models showed different levels of predictive success. Logistic Regression achieved a test F1-score of 0.70, reflecting its limitations in capturing complex, non-linear relationships in the dataset [4]. We used it for its simplicity and interpretability but it relatively struggled to model the finer interactions between features.

Random Forest turned out to be the best model in terms of performance, with a test F1-score of 0.85 and strong cross-validation results using a TimeSeriesSplit approach [5]. The model identified key predictors for inclusion, such as the 3-month average trading volume, number of analysts covering the stock, operating income, and book value per share. Conversely, predictors influencing removal or non-inclusion included the number of auditor changes and industries like Transportation & Public Utilities and Mining, potentially due to sufficient representation of these sectors in the index.

The Support Vector Classifier (SVC) underperformed relative to the other models, achieving a test F1-score of 0.74. Its results highlighted the model's sensitivity to feature scaling and its complexity in handling datasets with diverse feature distributions, making it less suitable for this problem.

The recall for class 0 (non-inclusion) was generally low across models - around 35-40% for SVC and Logistic Regression and a good ~68% for Random Forest basis which the prediction is not satisfactory for non-inclusion.

In terms of actual prediction for test data for 2023, we predicted the addition of Uber Technologies (UBER) and Jabil (JBL) starting the 3rd quarter but the actual addition happened in the 4th quarter. We rightly predicted the removal of SolarEdge Technologies (SEDG). Conversely, with this approach we missed the addition of BuildersFirstSource (BLDR) and removal of Alaska Air Group (ALK) and Sealed Air (SEE) to name a few.

## V. DISCUSSION

The results demonstrate the efficacy of using machine learning for financial forecasting. Random Forest emerged as the most effective model due to its ability to capture complex relationships among features. However, challenges such as overfitting and multicollinearity were addressed through rigorous feature engineering and validation strategies. The findings align with prior studies on the impact of strong financial viability on index membership.

Subjectively, the inclusion of non-traditional predictors, such as analyst coverage and restatement frequency, reflects a broader perspective on market behavior, emphasizing qualitative indicators' growing relevance. These results also underline the importance of transparency in such machine learning applications, as SHAP values highlighted feature contributions to predictions. Explainability was the key reason we chose such models and avoided use of black-box deep learning models to conduct a preliminary analysis.

A significant challenge encountered was the limited number of data points available for prediction. The initial approach focused on leveraging monthly financial and S&P data to capture recent trends and incorporate them into the predictive model. However, integrating monthly data from Compustat with S&P data resulted in approximately 85% missing values due to noise in the WRDS dataset.

While the reduced quarterly dataset provided strong cross-validated and out-of-sample results, the smaller sample size likely contributed to overfitting, particularly for predicting company additions to the S&P index. This highlighted the trade-off between data granularity and reliability, emphasizing the need for robust imputation or alternative data sources in future iterations. Exploration of using monthly data to predict quarterly changes leveraging other data sources is in our future scope of work.

## VI. FUTURE WORK

Based on the comprehensive analysis in this paper, future research and extensions of this work could focus on several promising directions. First, addressing the data granularity challenge by applying advanced imputation techniques or exploring alternative data sources could enhance model performance. A potential avenue would be integrating alternative financial databases, such as Bloomberg or FactSet, to mitigate the data completeness issues encountered with WRDS. Additionally, expanding the feature space to incorporate more nuanced predictors like ESG (Environmental, Social, and Governance) scores, social media sentiment analysis, and real-time economic indicators could provide a more holistic view of factors influencing S&P index membership. The research could also benefit from employing ensemble methods that combine multiple machine learning algorithms, potentially improving predictive accuracy and robustness [7]. Furthermore, developing a time-series approach that captures dynamic feature importance and tracks how predictive factors evolve could offer deeper insights into the mechanisms governing index membership.

These advancements could extend the model's applications to portfolio optimization, risk assessment, and policy analysis, driving deeper insights into market behavior.

## VII. SUMMARY

Our study successfully demonstrated a replicable approach to predicting S&P 500 membership changes using machine learning. The analysis highlighted key financial, market, and governance features as significant predictors while showcasing the potential of Random Forest models for such tasks. The ability to predict these changes holds a great alpha capture opportunity as discussed earlier. By addressing the challenges of data leakage and multicollinearity, the study ensures robust and actionable findings.

## VIII. ACKNOWLEDGEMENT

We would like to thank WRDS for providing access to high-quality datasets and S&P Dow Jones Indices for their detailed methodology documentation. Special thanks to all co-authors for collaborative efforts in developing the model, and to Professor Naftali Cohen for his invaluable guidance.